\documentclass[epj,nopacs]{svjour}
\usepackage{amsmath}
\usepackage{amssymb}
\usepackage{latexsym}
\usepackage[dvips]{graphicx}
\usepackage[english]{babel}
\usepackage{mathtools}

\def\bge{\begin{equation}}
\def\ene{\end{equation}}
\def\bgea{\begin{eqnarray}}
\def\enea{\end{eqnarray}}

\def\bge{\begin{equation}}
\def\ene{\end{equation}}
\def\bgea{\begin{eqnarray}}
\def\enea{\end{eqnarray}}

\def\ls{\raise 1.5pt\hbox{$\,<\;$}\kern -10.5pt\lower3.5pt
          \hbox{$\sim$}\kern 1.5pt} 
\def\gs{\raise 1.5pt\hbox{$\,>\,$}\kern -9.5pt\lower3.5pt
          \hbox{$\sim$}\kern 1.5pt} 

\usepackage{color}

\usepackage{ulem} 

\begin{document}
\sloppy
\title{A comparison of the $R_{\rm h}=ct$ and $\Lambda$CDM cosmologies \\
based on the observed halo mass function}

\author{Manoj K. Yennapureddy$^1$ and Fulvio Melia$^2$\thanks{John Woodruff Simpson Fellow.}}
\institute{$^1$Department of Physics, The University of Arizona, Tucson, AZ 85721, \email{manojy@email.arizona.edu}\\
           $^2$Department of Physics, the Applied Math Program, and Department of Astronomy,\\
           \null\hskip 0.1in The University of Arizona, Tucson, AZ 85721, \email{fmelia@email.arizona.edu}}

\authorrunning{Yennapureddy \& Melia}
\titlerunning{The Observed Halo Mass Function}

\date{June 5, 2019}

\abstract{The growth of structure may be traced via the redshift-dependent halo mass function.
        This quantity probes the re-ionization history and quasar abundance in the Universe,
        constituting an important probe of the cosmological predictions. Halos are
        not directly observable, however, so their mass and evolution must be inferred
        indirectly. The most common approach is to presume a relationship with galaxies and halos.
        Studies based on the assumption of a constant halo to stellar mass ratio
        $M_h/M_*$ (extrapolated from $z<4$) reveal significant tension with $\Lambda$CDM---a failure
        known as ``The Impossibly Early Galaxy Problem". But whether this ratio evolves or remains
        constant through redshift $4<z<10$ is still being debated. To eliminate the tension with
        $\Lambda$CDM, it would have to change by about $0.8$ dex over this range, an issue that 
        may be settled by upcoming observations with the James Webb Space Telescope. In 
        this paper, we explore the possibility that this major inconsistency may instead be an 
        indication that the cosmological model is not completely correct. We study this problem 
        in the context of another Friedmann-Lema\^itre-Robertson-Walker (FLRW) model known
        as the $R_{\rm h}=ct$ universe, and use our previous measurement of $\sigma_8$
        from the cosmological growth rate, together with new solutions to the
        Einstein-Boltzmann equations, to interpret these recent halo measurements. We
        demonstrate that the predicted mass and redshift dependence of the halo
        distribution in $R_{\rm h}=ct$ is consistent with the data, even
        assuming a constant $M_h/M_*$ throughout the observed redshift range
        ($4\lesssim z\lesssim 10$), contrasting sharply with the tension in $\Lambda$CDM.
        We conclude that---if $M_h/M_*$ turns out to be constant---the massive
        galaxies and their halos must have formed earlier than is possible in $\Lambda$CDM.}
\maketitle

\section{Introduction}
        The central principle behind the theory of structure formation is that large-scale
        assemblies, such as galaxies and clusters, formed via the growth of gravitational
        instabilities in the primordial density field, comprised of dark matter, radiation
        and baryonic matter. By assumption, dark matter is weakly interacting, so it
        decoupled from the radiation quite early and its fluctuations grew gravitationally
        to form the halos. Baryonic matter subsequently accreted into these potential wells
        once it also decoupled from the radiation, forming bound objects that would become
        stars and galaxies. Although this latter process is not yet fully understood, there is
        better consensus concerning the halo evolution itself, codified through the so-called
        halo mass function \cite{Sheth2001,Springel2005,Vogelsberger2014}.
        As it turns out, the halo mass function is highly sensitive to the
        cosmological parameters in $\Lambda$CDM, including the mass fraction $\Omega_{\rm m}$,
        the dark energy equation of state parameter $w_{\rm de}$, and $\sigma_8$
        \cite{Holder2001}---at least at lower redshifts (i.e., $z\leq 2$), where it plays a vital role
        in constraining standard cosmology. At higher redshifts, the halo mass function plays a
        vital role in probing the re-ionization history of the Universe \cite{Furlanetto2006}
        and the quasar abundance and formation sites \cite{HaimanLoeb2001}.
        It goes without saying that constraining and evaluating the halo mass function is
        therefore critical to the evaluation of structure formation in the Universe.

        The standard model predicts a rapid evolution in the number density of
        massive halos throughout the redshift range $10\gtrsim z\gtrsim 4$. If a strong connection
        exists between the halos and galaxies they host, one should expect to see a comparably
        rapid evolution in the number density of galaxies via their luminosity and mass distributions, implying that one should see in $\Lambda$CDM a sharp decline in the number density of luminous galaxies at constant luminosity, or a rapid decrease in luminosity for a fixed number density, towards high redshifts. Although the halo mass function has been evaluated using numerical and N-body simulations at high redshifts, only recently has it been tested observationally. The halos themselves are not directly observable, so they must be probed indirectly, e.g., through the measured galaxy mass distribution assuming a close relationship between the two.

        Data gathered recently with the Cosmic Assembly Near-Infrared Deep Extragalactic
        Survey (CANDELS; \cite{Grogin2011}), and the Spitzer Large Area Surveys (SPLASH; \cite{Capak2013}), allow us to now infer the halo mass function and its evolution with redshift. Past analyses of the halo mass function and galaxy luminosity function from these surveys have generated significant tension between the observations and predictions in the context of $\Lambda$CDM
\cite{Steinhardt2016}. These studies derived the halo mass from the UV luminosity function by assuming a relationship between UV luminosity and stellar mass which was then used to infer the halo mass function assuming another relationship between the stellar mass and halo mass. The outcome of this work \cite{Behroozi2013a} indicated that the halo to stellar mass ratio is constant throughout the redshift range $0<z<4$, but it is not yet clear whether this ratio evolves or remains constant at $z>4$.

In their analysis, Behroozi et al. \cite{Behroozi2013b}, Behroozi \& Silk \cite{BehrooziSilk2015} and
Finkelstein et al. \cite{Finkelstein2015} concluded that in order to alleviate the tension with $\Lambda$CDM,
this ratio needs to evolve by
as much as $\sim 0.8$ dex. An opposing view \cite{Steinhardt2016,RodriguezPuebla2017,Stefanon2017} maintains that such an evolution is not supported by existing data, and that this ratio is instead roughly constant at redshifts $4<z<10$, continuing the trend seen at $z<4$. In this case, the halo distribution would be inconsistent with $\Lambda$CDM by at least 2 to 4 orders of magnitude at redshifts $4<z<10$, a disparity termed as ``The Impossibly Early Galaxy Problem" \cite{Steinhardt2016}. It is anticipated that future observations with the NIRCam and NIRspec on the James Webb Space Telescope may settle this debate.

        In this paper, we consider what would happen if this problem turns out to be
        real and provide a possible solution using a recently completed study of the perturbation growth in the $R_{\rm h}=ct$ universe \cite{YennapureddyMelia2019} to describe and report the growth of structure from redshift $z\sim 10^{11}$ to $0$ in this alternative Friedmann-Lema\^itre-Robertson-Walker (FLRW) cosmology. We shall summarize the essential features
of this alternative model in \S~2, and then derive the growth equations in \S~2. We shall evaluate the halo mass
function in \S\S~3 and 4, and end with our conclusions in \S~5. 

\section{The $R_{\rm  h}=ct$ Universe}
The $R_{\rm h}=ct$ universe \cite{Melia2007,Melia2013a,Melia2016a,Melia2016b,MeliaAbdelqader2009,MeliaShevchuk2012} 
has thus far been tested using a variety of observations. For a summary, see Table~2 in ref.~\cite{Melia2018}. The principal 
difference between $\Lambda$CDM and $R_{\rm h}=ct$ is that the latter model is constrained by the equation of state $\rho+3p=0$, 
i.e., the so-called zero active mass condition in general relativity. In the standard model, radiation was dominant early, 
followed by matter and dark energy at later times, whereas dark energy has always been present in $R_{\rm h}=ct$, with a 
significant component of radiation early on, followed by matter towards lower redshifts. Also, the dark-energy equation 
of state is $w_{\rm de}=-1$ in $\Lambda$CDM, while it is $w_{\rm de}=-1/2$ in $R_{\rm h}=ct$ (see 
ref.~\cite{MeliaFatuzzo2016} for further details).

Some additional support for the $R_{\rm h}=ct$ cosmology, based on an alternative 
theoretical concept, may also be found in ref.~\cite{John:2000} and the updated 
discussion in ref.~\cite{John:2019}. But in spite of the success this model has enjoyed 
thus far in accounting for many observations as well, if not better, than the standard
model, some counter claims have also been published in recent years, so the issue
of whether or not it is the correct cosmology still needs to be resolved. This
is a principal reason for continuing to test it as we do in this paper. Our
analysis here advances this discussion significantly by providing new insights
and an important new comparison between $R_{\rm h}=ct$ and $\Lambda$CDM
using observations over an unusually large redshift range (see figs.~2-7 below).

As noted above, over the past decade, $R_{\rm h}=ct$ has been compared to $\Lambda$CDM 
using data across a broad redshift range, using integrated distances and the
redshift-dependent Hubble parameter, among various other measures. Still,
some of these data are often associated with unknown systematics and, worse, 
are often dependent on the presumed background model. The analysis of Type Ia
SNe is a well-known example in which the lightcurve is characterized by at
least 3 `nuisance' parameters that need to be optimized along with those in the
cosmological model. A different choice of assumptions (e.g., the unknown 
intrinsic dispersions) and techniques (e.g., $\chi^2$ minimization versus 
maximization of a likelihood function and/or model selection with information 
criteria), can sometimes produce varying outcomes in these tests.

For example, Type Ia SNe are challenging to use for model testing when various 
subsamples are merged together to improve the statistics, since one must deal
with different unknown systematics in each case. In his assessment, Shafer 
\cite{Shafer2015} merged the Union2.1 and JLA samples and found that this
compilation favours the standard model. In his analysis, however, he avoided 
the unknown intrinsic dispersions by instead constraining the reduced $\chi^2$
to be 1 in each subsample. In recent years, a superior statistical approach 
has been developed \cite{Kim2011,Wei2015,Melia2018e} in which these unknowns
are instead estimated by maximizing the overall likelihood function. The
outcome of which cosmology is preferred by the SN data changes depending
on which of these assumptions and methods are chosen.

Another recent test \cite{Lin2018} used local probes, combining SN data with 
measurements of the Hubble parameter $H(z)$ and baryon acoustic oscillations 
(BAO). This analysis also showed that $\Lambda$CDM is favoured over $R_{\rm h}=ct$,
contrasting with other work where the opposite was reported \cite{MeliaLopez2017,Melia2018e}.
The difference results may be traced to the choice of data sets in the two studies. 
As is the case for SN measurements, the BAO also do not provide model-independent 
information since the location of the BAO peak cannot easily be distinguished from 
redshift space distortions (RSD). As of today, only 3 such measurements have provided
a clean peak location. In other cases, a cosmology must be preassumed in order
to model the RSD, rendering the data highly model-dependent. Any use of these
BAO data, and the of $H(z)$ measured from them, produces a biased outcome. 
In their assessment, Lin et al. \cite{Lin2018} used all the data and concluded,
not surprisingly, that they favour $\Lambda$CDM because the standard model
was used to estimate the RSD. When only model-independent data are used 
instead, however, one reaches the opposite conclusion \cite{MeliaLopez2017}.

As we discuss elsewhere in this paper, the inferred halo mass function is itself 
subject to an important unknown: the redshift dependence of the halo to stellar 
mass ratio $M_h/M_*$, so our conclusions may also require revision once new data
will have been acquired. Depending on whether or not this ratio changes by roughly
an order of magnitude between redshifts 4 and 10, a factor yet to be resolved 
observationally, $\Lambda$CDM may or may not be favoured over $R_{\rm  h}=ct$. 
Nonetheless, providing one more important comparison between these two models 
is essential in establishing the conditions that must be met in order for 
$R_{\rm h}=ct$ to be viewed as a viable alternative to the standard model.

        \section{The Einstein-Boltzmann Equations for Dark Matter and Energy Fluctuations}
        To obtain the complete evolution of density fields starting from initial perturbations, one must solve the
        coupled Boltzmann/Einstein equations (see ref.~\cite{MaBertschinger1995}.
        The baryons are strongly coupled with radiation until decoupling and therefore do not
        contribute to the growth of structure during this epoch. Once they decouple from the
        radiation, baryons follow the evolution of dark matter, which has preceded them in
        forming bound systems. Hence, the initial growth of structure is dominated by dark
        matter. For this paper, which is focused on the question of halo growth, we
        therefore concentrate solely on the growth of dark matter perturbations. Other
        aspects of structure growth will be presented elsewhere \cite{YennapureddyMelia2019}. Thus,
        since we are not interested here in temperature fluctuations of
        the radiation field or acoustic oscillations, we justify the use of
        Einstein-Boltzmann equations customized solely for the purpose of describing
        the growth of dark matter perturbations, which we derive as follows.

The distribution function for any species (i.e., dark matter, baryons, radiation, etc.) may
depend on the coordinates ($x^\mu$) and momentum ($P^\mu$) 4-vectors, resulting in an
8-dimensional phase space. An additional constraint emerges, however, from the invariant
contraction of the momentum, $g_{\mu\nu}P^\mu P^\nu=-m^2$, which reduces the phase space to
7-dimensions. So we choose $x^\mu$, $p\equiv |\vec{p}|$ and the direction of the momentum,
$\hat{p}^i$, as our independent variables. Louisville's theorem produces the equation
         \begin{equation}
         \frac{df_s}{d\lambda}=\frac{\partial f_s}{\partial x^0}\frac{\partial x^0}{\partial \lambda}+
         \frac{\partial f_s}{\partial x^i}\frac{\partial x^i}{\partial \lambda}+
         \frac{\partial f_s}{\partial p}\frac{\partial p}{\partial \lambda}+
         \frac{\partial f_s}{\partial \hat{p}^i}\frac{\partial \hat{p}^i}{\partial \lambda}=C[f_s]\;,
         \end{equation}
where $f_s(x^\mu,p,\hat{p}^i)$ is the distribution function for any species `s' (i.e., dark matter,
dark energy, baryons, etc.), $\lambda$ is the affine parameter, and $C[f_s]$ is a collision/source
term for this species. We define $P^\mu={dx^\mu}/{d\lambda}$, so that ${dx^0}/{d\lambda}=P^0$.
Dividing the above equation by $P^0$, and neglecting the fourth term that is of second order,
gives
         \begin{equation}
         \frac{df_s}{d\eta}=\frac{\partial f_s}{\partial \eta}+\frac{\partial f_s}{\partial x^i}
         \frac{P^i}{P^0}+\frac{\partial f_s}{\partial p}\frac{\partial p}{\partial \eta}=
             \frac{C[f_s]}{P^0}\;,
         \end{equation}
where $\eta$ is now the conformal time, $d\eta\equiv dt/a(t)$, in terms of the expansion factor
$a(t)$ and cosmic time $t$ in the Friedmann-Lema\^itre-Robertson-Walker metric.

In the above equation, we may write ${P^i}/{P^0}=({p}/{E})\hat{p}^i$ (where $E$ is the energy) and,
using the geodesic equation, we get
         \begin{equation}
         \frac{dp}{d\eta}=-\mathcal{H}p+E\hat{p}^l\partial_l\frac{h_{00}}{2}-\frac{p}{2}
         \frac{dh_{ij}}{d\eta}\hat{p}^i\hat{p}^j\;,
         \end{equation}
where $\mathcal{H}$ is the Hubble parameter written in terms of $\eta$, and $h_{\alpha\beta}$
are the perturbed metric coefficients. Substituting Equation~(3) into Equation~(2),
we get
         \begin{eqnarray}
         \frac{df_s}{d\eta}&+&\frac{p\hat{p}^i}{E}\frac{\partial f_s}{\partial x^i}+p\bigg(-\mathcal{H}+
         \frac{E}{p}\hat{p}^l\partial_l\frac{h_{00}}{2}-\frac{1}{2}h_{ij}^{'}\hat{p}^i\hat{p}^j\bigg)
         \frac{\partial f_s}{\partial p}=\nonumber\\
         &\null&\qquad\qquad \frac{a}{E}(1-\Phi)C[f_s]\;,
         \end{eqnarray}
where we have substituted $P_0=\frac{E}{a}(1+\Phi)$, in terms of the perturbed gravitational
potential $\Phi$. We now separate the distribution function into its unperturbed component,
$\bar{f}_s$, and the perturbed contribution, $\mathcal{F}_s$, such that
         \begin{equation}
         f_s(\eta, x^i, p, \hat{p}^i)=\bar{f}_s(\eta, x^i, p, \hat{p}^i)+
             \mathcal{F}_s(\eta, x^i, p, \hat{p}^i)\;.
         \end{equation}
Then, multiplying Equation~(4) by $E(p)$, and integrating over momentum space, collecting the
zeroth-order terms, gives
         \begin{equation}
         \int \frac{d^3p}{(2\pi)^3}E(p)\frac{d\bar{f}_s}{d\eta}-\int \frac{d^3p}{(2\pi)^3}
            \mathcal{H}pE(p)
         \frac{\partial \bar{f}_s}{\partial p}=\int \frac{d^3p}{(2\pi)^3}aC[f_s]\;.
         \end{equation}
In this expression, $C[f_s]$ is zero in the context of $\Lambda$CDM because the particle number is
conserved during this phase of the fluctuation growth. But this is not the case in $R_{\rm h}=ct$.
The early universe in this model contains approximately $80\%$ dark energy and approximately
$20\%$ radiation, with a small contamination of matter \cite{MeliaFatuzzo2016}. At late times,
the $R_{\rm h}=ct$ universe contains approximately $70\%$ dark energy and $30\%$ matter. A coupling
therefore exists between dark matter and dark energy, such that the particle number for each
individual species is not conserved in this model. The right-hand side of Equation~(6) is
therefore not zero in $R_{\rm h}=ct$. Integrating the second term on the left-hand side by parts
and neglecting the boundary term, we arrive at the expression
         \begin{eqnarray}
         \int \frac{d^3p}{(2\pi)^3}E(p)\frac{d\bar{f}_s}{d\eta}&+&3\mathcal{H}\int\frac{d^3p}{(2\pi)^3}
         \bigg(E+\frac{p^2}{3E}\bigg)\bar{f}_s=\nonumber\\
         &\null&\qquad\qquad \int \frac{d^3p}{(2\pi)^3}aC[f_s]\;.
         \end{eqnarray}

This equation may be further reduced by using the following definitions for the (background)
density and pressure:
         \begin{equation}
         \rho_s=\int \frac{d^3p}{(2\pi)^3}E(p)\bar{f}_s\;,
         \end{equation}
and
         \begin{equation}
         \mathcal{P}_s=\int\frac{d^3p}{(2\pi)^3}\frac{p^2}{3E}\bar{f}_s\;.
         \end{equation}
When applied to dark matter, Equation~(7) may thus be written as follows
         \begin{equation}
         \frac{d\rho_{\rm dm}}{d\eta}+3\mathcal{H}(\rho_{\rm dm}+\mathcal{P}_{\rm dm})=
            \int \frac{d^3p}{(2\pi)^3}aC[f_{\rm dm}]\;.
         \end{equation}
For this particular species (i.e., dark matter), we may also put $\mathcal{P}_{\rm dm}=0$. In
addition, we use an approximate empirical expression, $\rho_{\rm dm}=({\rho_c}/{3a^2})
\exp\bigg(-\frac{a_*}{a}\frac{(1-a)}{(1-a_*)}\bigg)$, to model the transition from a
radiation/dark-energy dominated early universe to a matter/dark-energy dominated universe
at late times, where $\rho_c$ is the critical density today, and $a_*$ represents the
scale factor at matter radiation equality. Note that we are also normalizing $a(t_0)$ to
be $1$ today, which is possible in a spatially flat Universe. We infer the required
collision/source term in Equation~(10) by using this empirical expression for $\rho_{\rm dm}$,
which yields
         \begin{equation}
         \int \frac{d^3p}{(2\pi)^3}aC[f_{\rm dm}]=\mathcal{H}\rho_{\rm dm}+\frac{H\rho_{\rm dm}}{a}
            \bigg(\frac{a_*}{1-a_*}\bigg)\;.
         \end{equation}
It is not difficult to see that the above equation is satisfied to zeroth order only if
         \begin{equation}
         C[\bar{f}_{\rm dm}]=\frac{\mathcal{H}E}{a}\bar{f}_{\rm dm}+\frac{\mathcal{H}E}{a^2}
              \bar{f}_{\rm dm}\bigg(\frac{a_*}{1-a_*}\bigg)\;.
         \end{equation}
This collision/source term explicitly shows the interactions between dark energy and dark
matter required to sustain the zero active mass condition described above. That is, in order
for the partitioning of $80\%$ dark energy plus $20\%$ radiation in the early Universe to
transition to a balance of $70\%$ dark energy plus $30\%$ matter today, a fraction of the
dark energy must decay/evolve into dark matter. As such, the collision/source for dark
energy must be the negative of Equation~(12), so that
\begin{eqnarray}
\frac{d\rho_{\rm de}}{d\eta}&+&3\mathcal{H}(\rho_{\rm de}+\mathcal{P}_{\rm de})=\nonumber\\
&\null&\hskip-0.3in -\int\frac{d^3p}{(2\pi)^3}\bigg[\mathcal{H}E\bar{f}_{\rm dm}+\frac{\mathcal{H}E}{a}
\bar{f}_{\rm dm}\bigg(\frac{a_*}{1-a_*}\bigg)\bigg]\;,
\end{eqnarray}
where
\begin{equation}
C[\bar{f}_{\rm de}]=-\frac{\mathcal{H}E}{a}\bar{f}_{\rm dm}+\frac{\mathcal{H}E}{a^2}
\bar{f}_{\rm dm}\bigg(\frac{a_*}{1-a_*}\bigg)\;.
\end{equation}

Returning now to Equation~(4), and using Equation~(12), we find for dark matter that
         \begin{eqnarray}
         \frac{df_{\rm dm}}{d\eta}&+&\frac{p\hat{p}^i}{E}\frac{\partial f_{\rm dm}}{\partial x^i}+
            \bigg(-\mathcal{H}+\frac{E}{p}\hat{p}^l\partial_l\frac{h_{00}}{2}-\nonumber\\
            &\null&\hskip-0.7in \frac{1}{2}h_{ij}^{'}\hat{p}^i\hat{p}^j\bigg) \frac{\partial f_{\rm dm}}
            {\partial p}=
            \bigg[\mathcal{H}f_{\rm dm}+\frac{\mathcal{H}}{a}f_{\rm dm}
            \bigg(\frac{a_*}{1-a_*}\bigg)\bigg](1-\Phi)\;
         \end{eqnarray}
We again multiply this equation by $E(p)$ and integrate over momentum space, but now
collecting first order terms, finding that
         \begin{eqnarray}
         \hskip-0.1in \frac{d(\delta \rho_{\rm dm})}{d\eta}&+&(\rho_{\rm dm}+\mathcal{P}_{\rm dm})
            \partial_iv^i_{\rm dm}
            +3\mathcal{H}(\delta \rho_{\rm dm}+\delta \mathcal{P}_{\rm dm})+\nonumber\\
            &\null&\hskip-0.7in  3(\rho_{\rm dm}+ \mathcal{P}_{\rm dm})\frac{d\Phi}{d\eta}= 
            \bigg[\mathcal{H}+\frac{\mathcal{H}}{a}
            \bigg(\frac{a_*}{1-a_*}\bigg)\bigg](1-\Phi)\delta \rho_{\rm dm}\quad
         \end{eqnarray}
where, as always, $\Phi$ is the gravitational potential. Then, defining
$\delta_{\rm dm}\equiv{\delta\rho_{\rm dm}}/{\rho_{\rm dm}}$ for the dark-matter perturbation,
with $\mathcal{P}_{\rm dm}=\delta\mathcal{P}_{\rm dm}=0$, we may write
         \begin{equation}
         \frac{d\delta_{\rm dm}}{d\eta}=\frac{1}{\rho_{\rm dm}}\frac{d(\delta\rho_{\rm dm})}{d\eta}-
         \frac{\delta_{\rm dm}}{\rho_{\rm dm}}\frac{d\rho_{\rm dm}}{d\eta}\;.
         \end{equation}
Substituting for $d(\delta\rho_{\rm dm})/d\eta$ in Equation~(16), and isolating the
Fourier mode $k$, we find that
        \begin{equation}
        \frac{d\delta_{{\rm dm},k}}{d\eta} =-ku_k -3\frac{d\Phi_k}{d\eta}-\mathcal{H}
          \bigg[1+\frac{a_*}{a(1-a_*)}\bigg]\Phi\;,
        \end{equation}
where we have written $\partial_iv^i_k=ku_k$, in terms of the velocity perturbation $u_k$
of the dark matter. Finally, we take the second moment of Equation~(15),
multiplying it by $p\hat{p}^i$ and contracting it with $i\hat{k}_i$. Then integrating over
momentum space, and collecting first order terms, we get
\begin{eqnarray}
\frac{d(\rho_{\rm dm}u_{\rm dm,k})}{d\eta}&+&4\mathcal{H}\rho_{\rm dm}u_{\rm dm,k}+k\Phi\rho_{\rm dm}=\nonumber\\
&\null&\mathcal{H}\bigg[1+\frac{a_*}{a(1-a_*)}\bigg]\rho_{\rm dm}u_{\rm dm,k}\;,
\end{eqnarray}
where $u_{\rm dm,k}$ is the $k^{\rm th}$ velocity perturbation of dark matter. Substituting for
${d\rho_{\rm dm}}/{d\eta}$ in the above equation we thus get
        \begin{equation}
        \frac{du_{\rm dm,k}}{d\eta}=-\frac{1}{a}\frac{da}{d\eta}u_{\rm dm,k}-k\Phi_k\;.
        \end{equation}

Turning now to the dark-energy perturbations, we begin with Equation~(4) and the interaction term
in Equation~(14), and find that
\begin{eqnarray}
\frac{df_{\rm de}}{d\eta}&+&\frac{p\hat{p}^i}{E}
\frac{\partial f_{\rm de}}{\partial x^i}+p\bigg(-\mathcal{H}+
\frac{E}{p}\hat{p}^l\partial_l\frac{h_{00}}{2}-\nonumber\\
&\null&\hskip-0.7in\frac{1}{2}h_{ij}^{'}\hat{p}^i\hat{p}^j\bigg)
\frac{\partial f_{\rm de}}{\partial p}=
-\mathcal{H}(1-\Phi)\bigg[1+\frac{a_*}{a(1-a_*)}\bigg]f_{\rm dm}\;,
\end{eqnarray}
where $f_{\rm de}$ is the distribution function for dark-energy. Thus, partitioning $f_{\rm de}$ into
its unperturbed ($\bar{f_{\rm de}}$) and perturbed ($\mathcal{F}_{\rm de}$) components, as was done
in Equation~(5), we can collect the first-order perturbed terms to find that
\begin{eqnarray}
\frac{d\mathcal{F}_{\rm de}}{d\eta}&+&\frac{p\hat{p}^i}{E}\frac{\partial \mathcal{F}_{\rm de}}{\partial x^i}-
\mathcal{H}p\frac{\partial \mathcal{F}_{\rm de}}{\partial p}+p\bigg(\frac{E}{p}\hat{p}^l\partial_l
\frac{h_{00}}{2}-\nonumber\\
&\null&\hskip-0.5in\frac{1}{2}h_{ij}^{'}\hat{p}^i\hat{p}^j\bigg)\frac{\partial \bar{f}_{\rm de}}
{\partial p}= -\mathcal{H}\bigg[1+\frac{a_*}{a(1-a_*)}\bigg]\bigg(\mathcal{F}_{\rm dm}-
\Phi\bar{f}_{\rm dm}\bigg).\qquad
\end{eqnarray}
Multiplying this equation by $E(p)$ and integrating over the momentum space then gives
\begin{eqnarray}
\hskip-0.2in\frac{d \delta \rho_{\rm de}}{d\eta}&+&\rho_{\rm de}(1+w_{\rm de})ku_{{\rm de},k}+3\mathcal{H}
\delta \rho_{\rm de}\bigg(1+\frac{\delta \mathcal{P}_{\rm de}}{\delta \rho_{\rm de}}\bigg)+\nonumber\\
&\null&\hskip-0.6in 3 \frac{d\Phi}{d\eta}\rho_{\rm de}(1+w_{\rm de})=
\mathcal{H}\bigg[1+\frac{a_*}{a(1-a_*)}\bigg](\rho_{\rm dm}\Phi-\delta \rho_{\rm dm}).
\end{eqnarray}
Defining $\delta_{\rm de}={\delta \rho_{\rm de}}/{\rho_{\rm de}}$, and using
$\mathcal{P}_{\rm de}=-\rho_{\rm de}/2$ (Melia \& Fatuzzo 2016) we may write
\begin{equation}
\frac{d\delta_{\rm de}}{d\eta}=\frac{1}{\rho_{\rm de}}\frac{d \delta \rho_{\rm de}}{d\eta}-
\frac{\delta \rho_{\rm de}}{\rho_{\rm de}^2}\frac{d\rho_{\rm de}}{d\eta}\;,
\end{equation}
so that with Equations~(13) and (23), we find that
\begin{eqnarray}
\frac{d\delta_{\rm de}}{d\eta}&=&-\frac{k}{2}u_{{\rm de},k}-3\mathcal{H}\delta_{\rm de}
\bigg(\frac{1}{2}+\frac{\delta \mathcal{P}_{\rm de}}{\delta \rho_{\rm de}}\bigg)-
\frac{3}{2}\frac{d\Phi}{d\eta}+\nonumber\\
&\null& \mathcal{H}\bigg[1+\frac{a_*}{a(1-a_*)}\bigg]
\frac{\rho_{\rm dm}}{\rho_{\rm de}}(\delta_{\rm de}-\delta_{\rm dm}+\Phi)\;.
\end{eqnarray}

The sound speed for our coupled dark matter/dark energy fluid is not known yet, so
we write it as follows
\begin{equation}
c_s^2\equiv\frac{\delta \mathcal{P}}{\delta \rho}=\frac{\delta \mathcal{P}_{\rm de}}{\delta \rho_{\rm dm}+\delta 
\rho_{\rm de}}=\frac{\delta \mathcal{P}_{\rm de}/\delta \rho_{\rm de}}{(1+\delta \rho_{\rm dm}/\delta+
\rho_{\rm de})}\;,
\end{equation}
analogously to what is commonly done with the coupled baryon-radiation fluid in the
standard model. And following the conventional approach of assuming adiabatic fluctuations,
we also write
\begin{equation}
\frac{\delta \mathcal{P}_{\rm de}}{\delta \rho_{\rm de}}=c_s^2\bigg[1+\frac{2\rho_{\rm dm}}
{\rho_{\rm de}}\bigg]\;.
\end{equation}

For the sake of simplicity, we assume the sound speed to be a constant delimited to the range
$0<(c_s/c)^2<1$. We have found that the actual value of this constant has a negligible impact
on the solutions to the above equations since the ratio of dark matter density to dark energy
is always much less than $1$ in the $R_{\rm h}=ct$ universe, and we therefore adopt the simple
fraction $c_s^2=c^2/2$ throughout this work. Thus, using Equations~(25) and (27), we get
\begin{eqnarray}
\frac{d\delta_{\rm de}}{d\eta}&=&-\frac{k}{2}u_{{\rm de},k}-\delta_{\rm de}\bigg(
\frac{3\mathcal{H}}{2}+3\mathcal{H}c_s^2+\frac{6\mathcal{H}c_s^2\rho_{\rm dm}}
{\rho_{\rm de}}\bigg)-\nonumber\\
&\null&\hskip-0.4in\frac{3}{2}\frac{d\Phi}{d\eta}+ \mathcal{H}\bigg[1+
\frac{a_*}{a(1-a_*)}\bigg]\frac{\rho_{\rm dm}}{\rho_{\rm de}}(\delta_{\rm de}-\delta_{\rm dm}+\Phi)\;.
\end{eqnarray}
Finally, we take the second moment of Equation~(21), multiply it by $p\hat{p}^i$ and contract it
with $i\hat{k}_i$. Integrating over momentum space, and collecting first order terms, we thus find that
\begin{eqnarray}
\frac{du_{{\rm de},k}}{d\eta}&=&-\frac{5\mathcal{H}}{2}u_{\rm de,k}-k\Phi+2kc_s^2\bigg[1+\frac{2\rho_{\rm dm}}
{\rho_{\rm de}}\bigg]\delta_{\rm de}+ \nonumber\\
&\null&\hskip-0.2in\mathcal{H}\bigg[1+\frac{a_*}{a(1-a_*)}\bigg]\frac{\rho_{\rm dm}}
{\rho_{\rm de}}(u_{{\rm de},k}-2u_{{\rm dm},k})\;.
\end{eqnarray}

Our final equation comes from perturbing the FLRW metric in Einstein's Equations
(see Ma \& Bertschinger 1995), which gives
        \begin{eqnarray}
        k^2\Phi_k&+&3\frac{1}{a}\frac{da}{d\eta}\bigg(\frac{d\Phi_k}{d\eta}+
        \frac{1}{a}\frac{da}{d\eta}\Phi_k\bigg)=\nonumber\\
        &\null&\qquad\qquad\quad
        4\pi G a^2 \bigg[\rho_{\rm m} \delta_{{\rm dm},k}+\rho_{\rm de}\delta_{\rm de} \bigg]\;.
        \end{eqnarray}
In arriving at Equation~(30), we have chosen the Newtonian gauge for the primary reason that the independent
components in this gauge have a direct correspondence to the gauge invariant Bardeen variables
\cite{MaBertschinger1995,Bardeen1980}.

It is important to stress that the set of Equations~(18, 20, 28, 29) in $R_{\rm h}=ct$
differ from their counterparts in $\Lambda$CDM. This happens because dark energy and dark matter
are coupled in $R_{\rm h}=ct$, while dark energy is simply a cosmological constant in the most basic
$\Lambda CDM$ model. The only expression that is formally common to both $R_{\rm h}=ct$ and
$\Lambda CDM$ is Equation~(20), though the dependence
of $\rho_{\rm dm}$ on $a(t)$ is, of course, model dependent. Mathematically, this comes about
because the collision/source term in Equation~(12) actually cancels out in the perturbation
Equation~(19) for the velocity perturbations. The dependence on cosmology also enters
into the growth of $\delta_{\rm dm}$ via the
model-dependent $\mathcal{H}$ and $a(t)$ functions. These quantities change with time according to
the background evolution, and are therefore strongly dependent on the chosen model.

        \begin{figure}
                \centering
                \includegraphics[scale=0.55]{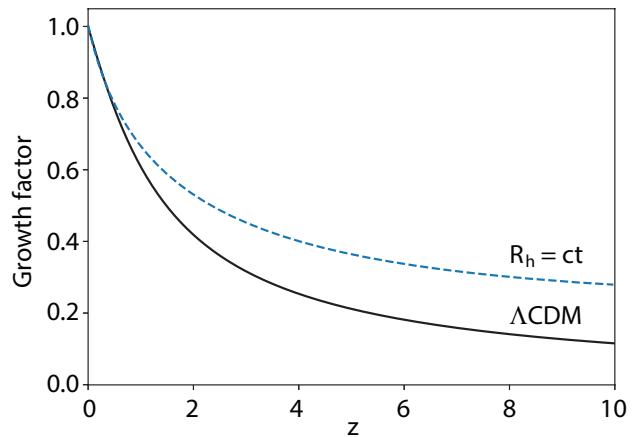}
                \label{Fig 1}\caption{Growth Factor predicted by $R_{\rm h}=ct$ (dashed)
                        and flat $\Lambda$CDM (solid).}
        \end{figure}

More specifically, an inspection of these equations reveals that there are three principal differences
between $R_{\rm h}=ct$ and $\Lambda$CDM: i) the scale factor $a(t)$ in $R_{\rm h}=ct$ is given as
$a(t)=(t/t_0)$ at all epochs, whereas in $\Lambda$CDM it is proportional to $t^{1/2}$ and $t^{2/3}$
during the radiation and matter dominated phases, respectively; ii) the matter density
scales as $\rho_{\rm dm}=({\rho_c}/{3a^2})\exp\bigg(-\frac{a_*}{a}\frac{(1-a)}{(1-a_*)}\bigg)$ in
$R_{\rm h}=ct$, whereas it is given as $\rho_{\rm m}=(\Omega_m\rho_c/a^3)$ in $\Lambda$CDM; and
iii) the various modes of the density field in $\Lambda$CDM exited the horizon during inflation,
whereas none of the modes ever crossed the horizon in $R_{\rm h}=ct$ \cite{Melia2017}. In $\Lambda$CDM,
small-scale modes re-entered the horizon while radiation was dominant, while larger-scale modes
entered the horizon when matter dominated, which produces a late start for the growth of
structure compared with what happens in $R_{\rm h}=ct$. This appears to be the principal
reason why galaxies and supermassive black holes appeared earlier in $R_{\rm h}=ct$
than in the standard model.

        We shall formally introduce the growth function in Equation~(31) below, and plot it
        in figure~1. It is obtained by solving Equations~(10--30) simultaneously (see
        ref.~\cite{YennapureddyMelia2019} for more details). It is quite evident from this plot that the
        growth factor in $R_{\rm h}=ct$ is significantly stronger at large redshifts than that
        in $\Lambda$CDM, in full agreement with the previous results of our analysis at lower
        redshifts \cite{YennapureddyMelia2018}. In contrast, the growth function
        in $\Lambda$CDM indicates a strong evolution from $z\sim 10$ to $z\sim 4$. And since
        galaxies typically form on a dynamical timescale $\sim 300$ Myr \cite{Wong2009} after
        halo virialization, the rapid evolution in the number density of halos from
        $z\sim 8$ to $\sim 4$ predicted by $\Lambda$CDM corresponds to a rapid evolution
        in the UV luminosity of galaxies at $6.0\gtrsim z\gtrsim 3.4$. This is one
        of the points of contention between the two camps, since this (required) rapid
        evolution in the UV luminosity function conflicts with the observations \cite{Steinhardt2016}.
        The observed UV luminosity evolves much more slowly than this prediction,
        which would mean that these massive galaxies would have formed much earlier than
        expected in $\Lambda$CDM.

        \begin{figure}
                \centering
                \includegraphics[scale=0.8]{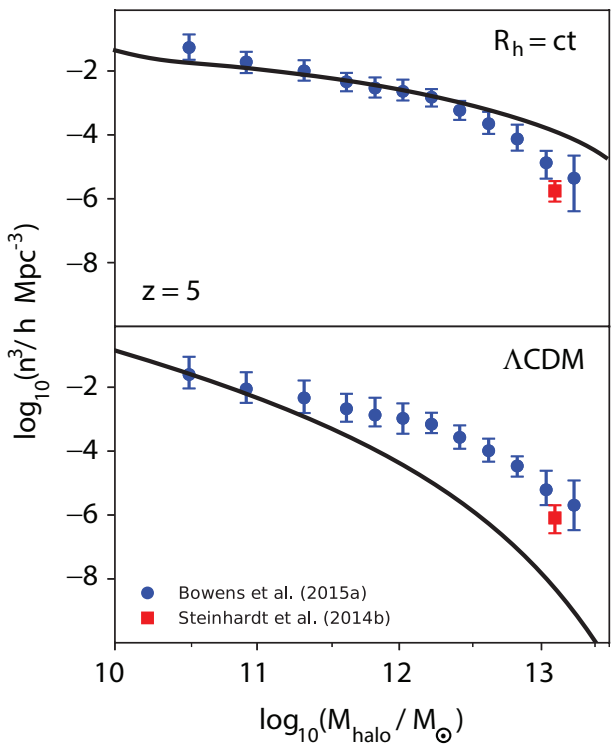}
                \label{Fig 2}\caption{Top: halo mass function inferred from galaxy surveys at $z=5$
                        compared with $R_{\rm h}=ct$. Bottom: Same, except now for $\Lambda$CDM.}
        \end{figure}

        \section{Halo Mass Function}
        The halo mass function was first derived by Press and Schechter \cite{PressSchechter1974}  assuming
        spherical collapse and a primordial Gaussian density field. When tested against numerical
        simulations, however, it became evident that the Press-Schechter formalism
        over-predicts the number of halos at the high mass end, and under-predicts at
        the low mass end. This inconsistency was resolved by introducing ellipsoidal
        collapse, rather than spherical, by Sheth-Tormen \cite{Sheth2001}. But the Bolshoi
        simulations performed by Klypin, Trujillo-Gomez \& Primack \cite{Klypin2011} several years later
        indicated that, while discrepancies in the Sheth-Tormen mass function at $z\sim 0$
        are less than $10\%$ for halo masses in the range $5\times10^{9} - 5\times10^{14}\;M_{\odot}$,
        this prescription over-predicts the density by about $50\%$ at $z\sim 6$ for masses
        in the range $10^{11}-10^{12}\;M_{\odot}$, getting even worse (by an order of
        magnitude) by $z\sim 10$. Unfortunately for the standard model, the inclusion
        of corrections from the Bolshoi simulations actually exacerbates the discrepancy
        between theory and observation. For this reason, and the fact that analogous
        simulations to the Bolshoi calculations have not yet been carried out for
        $R_{\rm h}=ct$, we won't include such adjustments in this paper. We
        point out that if we were to add such corrections to $R_{\rm h}=ct$, the
        comparison of this model's predictions with the data under the
        assumption of a constant halo mass to stellar mass ratio would be even more
        favourable than the Sheth-Tormen formulation on its own, as one may readily see in
        figures~2-7. As such, our exclusion of these corrections produces an
        effect more favourable to $\Lambda$CDM than $R_{\rm h}=ct$, even with this assumption,
        which we do in order give the standard model as much benefit of the doubt as possible.

        The Sheth-Tormen mass function is given as
        \begin{equation}
        f(\sigma)=A\sqrt{\frac{2a}{\pi}}\bigg[1+\bigg(\frac{\sigma ^2}{a\delta_c^2}\bigg)^p\bigg]
        \frac{\delta_c}{\sigma}\exp\bigg[-\frac{a\delta_c^2}{2\sigma^2}\bigg]\;,
        \end{equation}
        where $A=0.3222$ is a normalization factor, and $a=0.707$ and $p=0.3$. Using this halo mass
        function, one may obtain the number of dark matter halos per comoving volume with masses
        less than $M$ as follows:
        \begin{equation}
        \frac{dn}{d\ln M}=\frac{\rho_0}{M} f(\sigma)\bigg|\frac{d\ln\sigma}{d\ln M}\bigg|\;,
        \end{equation}
        where $\sigma$ is defined according to the expression
        \begin{equation}
        \sigma_R^2(R,z)=\frac{b^2(z)}{2\pi^2}\int_{0}^{\infty}k^2P(k)W^2(k,R)dk\;,
        \end{equation}
        and $P(k)$ is the power spectrum, $W(k,R)$ is the top-hat filter and $b(z)$ is the
        growth factor shown in figure~1 for both $\Lambda$CDM and $R_{\rm h}=ct$.

        \section{Observed Halo Mass Function}
        The data used in this paper were assembled by Steinhardt et al. \cite{Steinhardt2016},
        based on measurements obtained using three different techniques, including
        the clustering method \cite{Hildebrandt2009,Lee2012} based on the
        spatial distribution of galaxies to obtain the halo masses. This method doesn't
        assume any physical properties of the galaxies themselves, but assumes a model
        for the dark matter concentration. Other techniques include template fitting
        \cite{Ilbert2013}, that adopts a relationship between the luminosity and
        stellar masses; the abundance matching technique \cite{Finkelstein2015}
        that relates critical features in the galaxy luminosity or mass function,
        such as a `knee', to crucial elements in the halo mass distribution, that
        can then be used to match the galaxy and dark matter densities to infer the
        halo mass function. The high redshift ($z\ge6$) data points are derived from
        the UV luminosity function, that yields halo masses by assuming that the
        halo mass to light ratio obtained at lower redshifts persists to higher
        redshifts. Most of the data used in this work were obtained
        assuming a constant ratio of halo to stellar-mass. The two main principles
        for arriving at this ratio are (i) that $10\%$ of the baryonic matter
        eventually condensed into stars \cite{Leauthaud2012} and (ii) the
        observation of a 6:1 ratio of dark matter to baryonic matter \cite{Planck2016}.

        \begin{figure}
                \centering
                \includegraphics[scale=0.8]{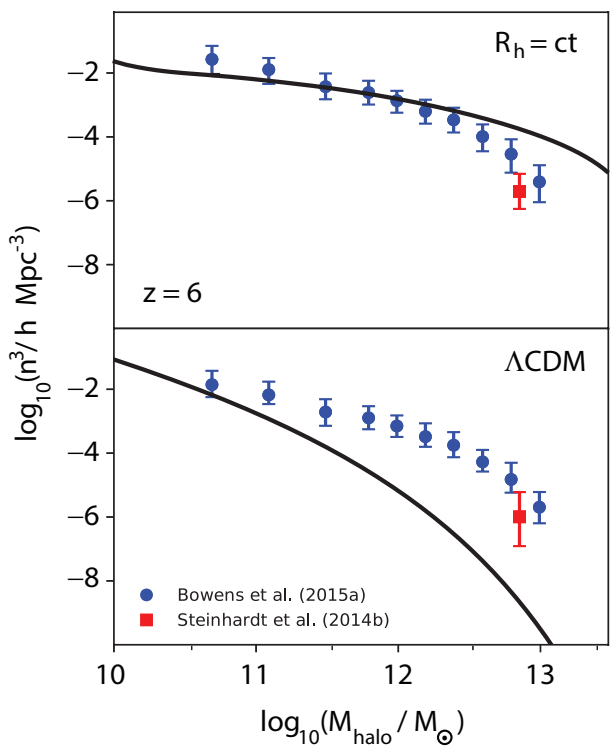}
                \label{Fig 3}\caption{Top: halo mass function inferred from galaxy surveys at $z=6$
                        compared with $R_{\rm h}=ct$. Bottom: Same, except now for $\Lambda$CDM.}
        \end{figure}

        \begin{figure}
                \centering
                \includegraphics[scale=0.8]{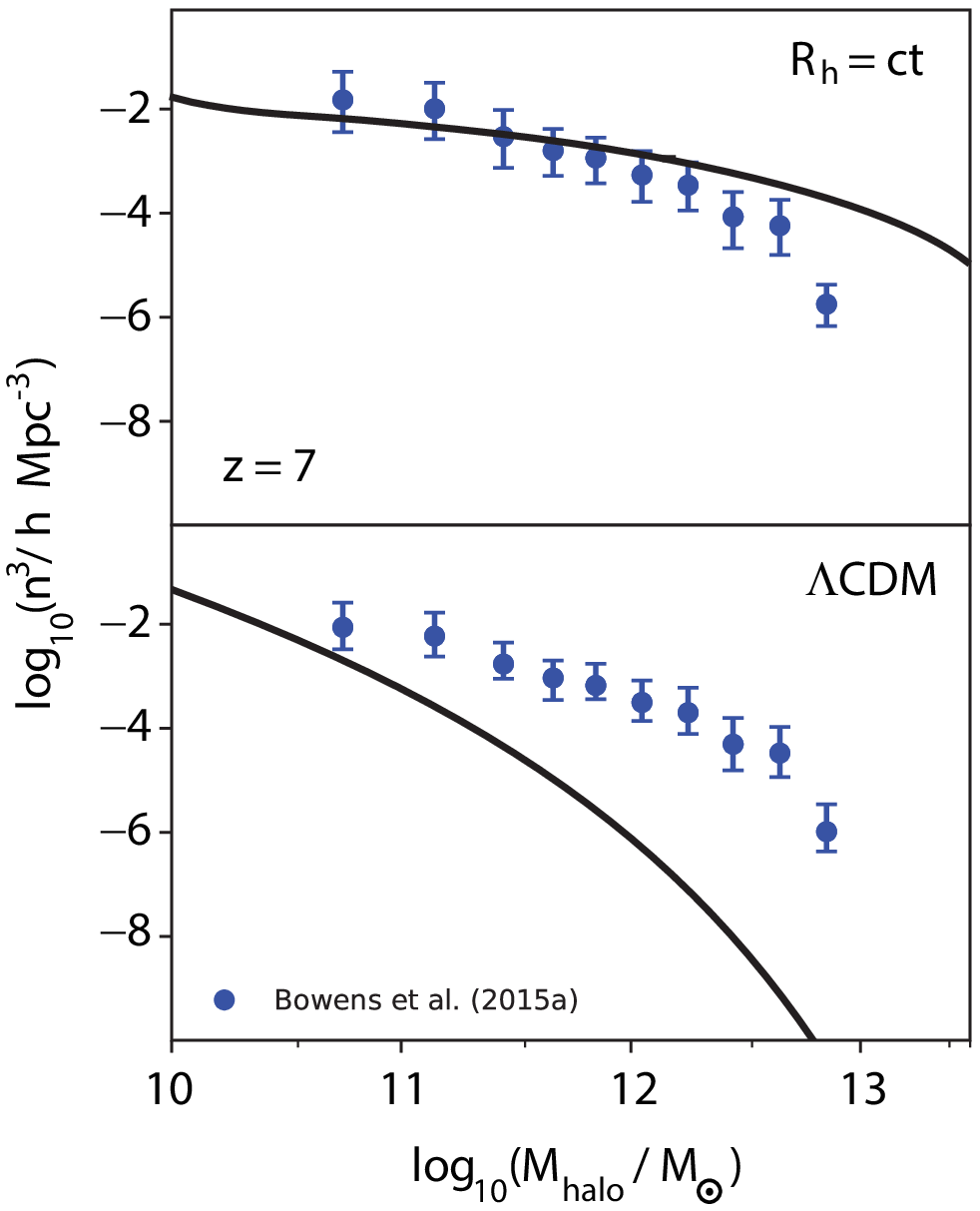}
                \label{Fig 4}\caption{Top: halo mass function inferred from galaxy surveys at $z=7$
                        compared with $R_{\rm h}=ct$. Bottom: Same, except now for $\Lambda$CDM.}
        \end{figure}

        \begin{figure}
                \centering
                \includegraphics[scale=0.8]{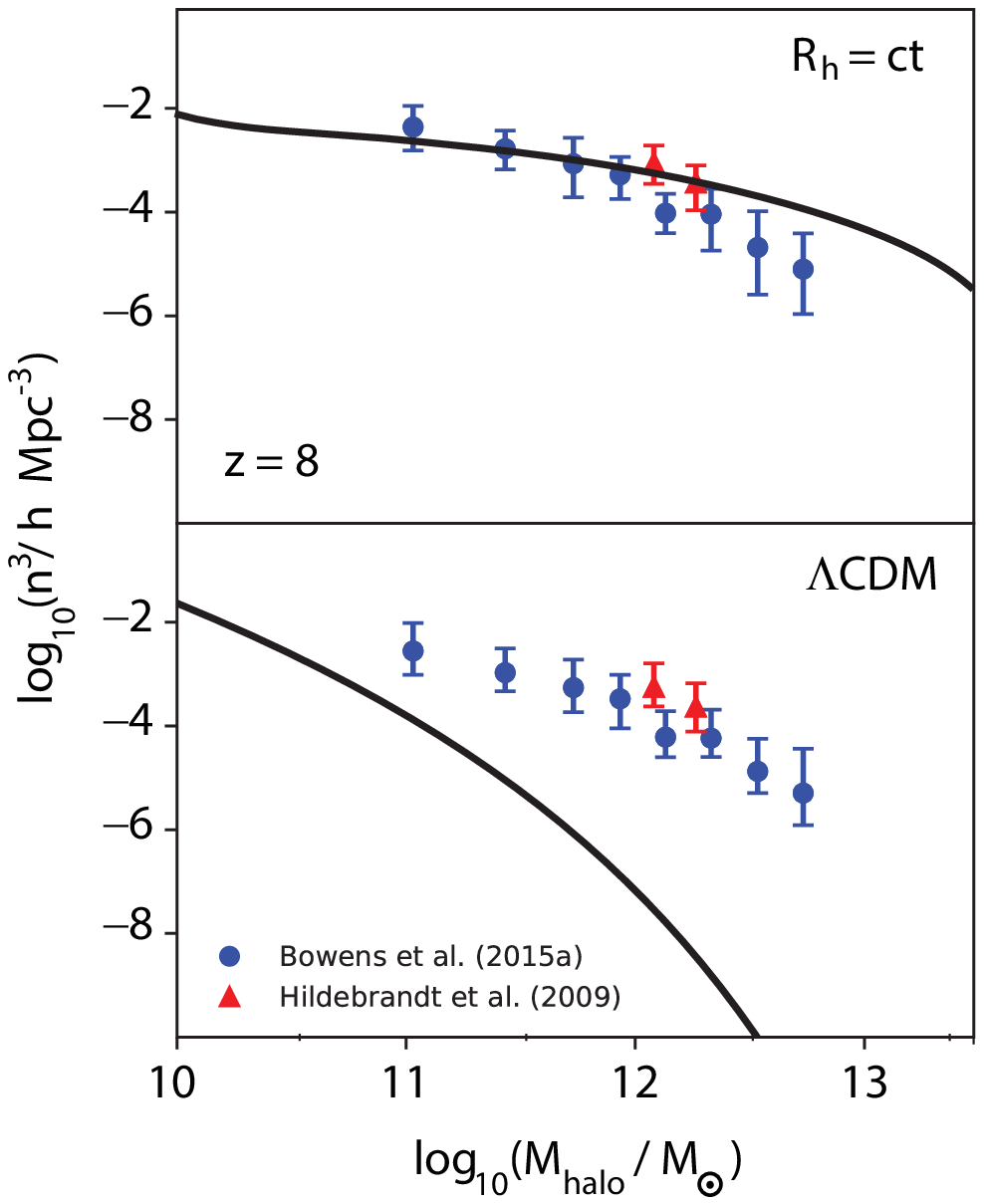}
                \label{Fig 5}\caption{Top: halo mass function inferred from galaxy surveys at $z=8$
                        compared with $R_{\rm h}=ct$. Bottom: Same, except now for $\Lambda$CDM.}
        \end{figure}

        \begin{figure}
                \centering
                \includegraphics[scale=0.8]{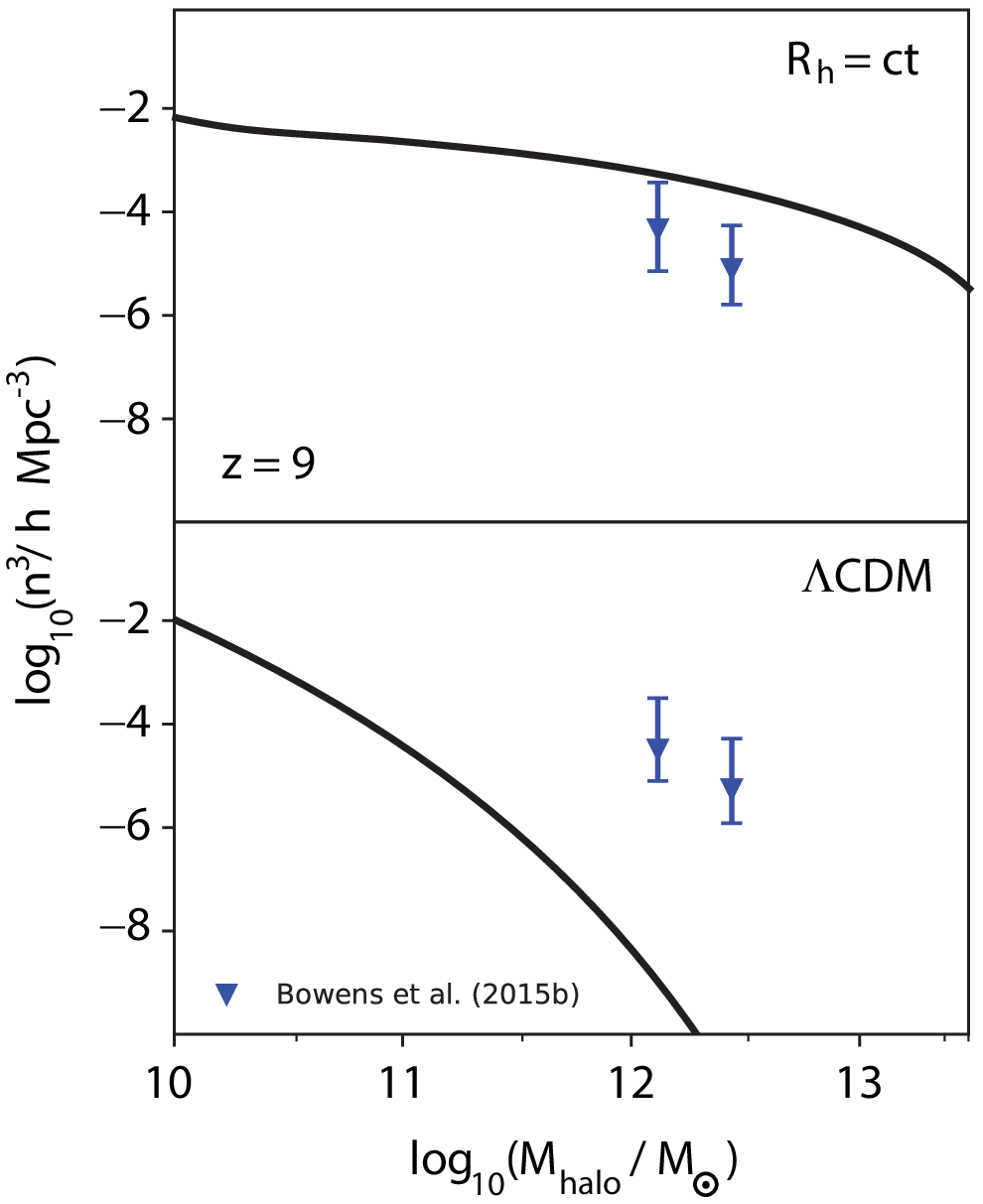}
                \label{Fig 6}\caption{Top: halo mass function inferred from galaxy surveys at $z=9$
                        compared with $R_{\rm h}=ct$. Bottom: Same, except now for $\Lambda$CDM.}
        \end{figure}

        \begin{figure}
                \centering
                \includegraphics[scale=0.8]{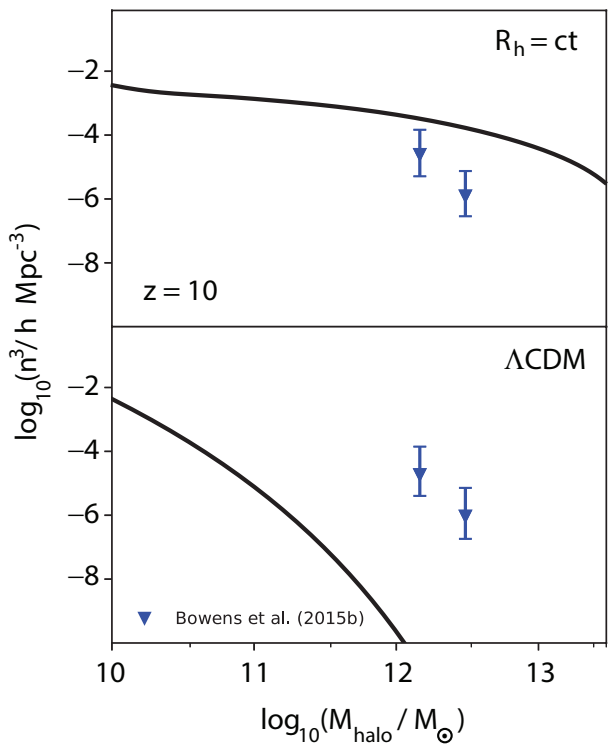}
                \label{Fig 7}\caption{Top: halo mass function inferred from galaxy surveys at $z=10$
                        compared with $R_{\rm h}=ct$. Bottom: Same, except now for $\Lambda$CDM.}
        \end{figure}

        It is quite obvious from the progression seen in figures~2--7 that the
        observed halo mass function obtained via these different techniques
        \cite{Hildebrandt2009,Lee2012,Finkelstein2015,Caputi2015} is entirely
        inconsistent with the distribution predicted by $\Lambda$CDM, if the halo to stellar-mass
        ratio remains constant throughout the $4<z<10$ redshift range \cite{Steinhardt2016}. 
        Of course, the caveat is that these data were not measured directly, and
        were obtained using relationships derived at low redshifts. Steinhardt et al. \cite{Steinhardt2016}
        studied the possibility that these correlations could be breaking down
        at high-$z$. Their investigation indicated, however, that the star-formation rate
        vs. stellar mass of these high redshift galaxies lies on the extrapolation from
        lower redshift galaxies. In addition, the ratio of stellar mass to halo
        mass in these high redshift galaxies is similar to the standard value 30:1
        seen at all redshifts. These two tests therefore indicate that the high redshift
        galaxies are quite normal, implying that the problem is real.

        In addition to this, Steinhardt et al. \cite{Steinhardt2016} determined that an evolution of
        0.8 dex in $M_{Halo}/L_{UV}$ is needed to mitigate this problem. Such a
        change might occur if the stellar population in galaxies at $z=8$ is younger
        than that at $z=4$. Steinhardt et al. \cite{Steinhardt2016} extensively investigated whether
        this possibility could mitigate the disparity by modeling the halo mass to light
        ratio from an initial stellar population assuming they formed in one rapid burst
        at $z=12$ and then evolved along the main sequence until $z=4-8$, where they
        were observed. This resulted in a star formation rate $\propto M_*^{0.7}$, with
        a stellar age asymptotically approaching $50-150$ Myr, starting from an initially
        small value. But this isn't sufficient enough to remove the problem and, worse,
        the above approach isn't realistic considering a dynamical timescale of
        $300$ Myr for star formation after virialization of the halo.

        Steinhardt et al. \cite{Steinhardt2016} considered this scenario and modeled the halo mass to
        light ratio as described above, concluding that this too is insufficient to
        reconcile the problem. Another possibility is that the halo mass to
        stellar mass ratio evolves towards higher redshifts. An evolution of 0.8 dex
        in this ratio would reconcile the problem. But such a modification is only
        possible either by a complete absence of dark matter at redshift $8$, or if
        $100\%$ of the baryons condensed instantly into stars at high redshift upon
        halo virialization, which is quite impossible. Hence, one may reasonably
        conclude that this problem may be reconciled in $\Lambda$CDM only via the
        introduction of implausible physics. When viewed in the context of other
        ``too early'' types of problems, the disparity evident in figures~2-7 is
        quite damning for the standard model. For example, the early appearance of
        supermassive black holes at $z\sim 6-7$ \cite{Melia2013b,MeliaMcClintock2015}
        and galaxies at $z\sim 10-12$ (see references cited in \cite{Melia2014}, argues in 
        favor of these problems being real, presenting a challenge to any attempt to 
        alleviate them in the context of $\Lambda$CDM.

In contrast, the comparison between the Steinhardt et al. \cite{Steinhardt2016} data, under
the assumption that the halo to stellar mass ratio is constant in the redshift range $4<z<10$, and the
predictions of $R_{\rm h}=ct$, is very favourable---except at the very high mass end of the halo mass
distribution, as one may see in Figures~2-7. The standard model disagrees progressively more and more
with this approach as the redshift increases, while $R_{\rm h}=ct$ fits the data throughout the range
$10\gtrsim z\gtrsim 4$ very well at the low and intermediate mass end, and overpredicts by one to two
orders of magnitude at the high mass end. This over-prediction may be due to two possible reasons: (1) As
noted earlier, the Bolshoi simulation \cite{Klypin2011} has indicated that the Sheth-Tormen mass function
overpredicts the number of halos by at least 10\% at redshift $z\sim 0$, and overpredicts by at least
50\% at redshift $z \sim 10$. Although simulations similar to Bolshoi haven't yet been carried out
for $R_{\rm h}=ct$, a trend analogous to this in the context of this model, would produce corrections
that largely mitigate the problem at the high mass end; (2) This over-prediction may also be due in
part to observational selection effects that may be `hiding' some of the sources. Some massive galaxies
may have been missed due to extinction, which future observations might be able to address. Regardless
of which, if any, of these mitigating factors are at play in $R_{\rm h}=ct$, none of them can resolve the
disparity arising from the predictions of $\Lambda$CDM. The discrepancy seen in the standard model is
extreme, ranging from one to over four orders of magnitude from low to high mass, throughout the redshift
range $4\lesssim z\lesssim 10$. The factors that may alleviate the high-mass end problem with $R_{\rm h}=ct$,
actually makes the comparison much worse for $\Lambda$CDM, increasing the disparity between predictions
and observations. The weaker evolution in growth rate predicted by $R_{\rm h}=ct$ is the vital reason for
its success, indicating that massive galaxies must have formed earlier than predicted in the standard model,
consistent with the observations.

        The problem in $\Lambda$CDM may instead be reconciled with an evolution in the halo
        mass to light ratio, which could happen, e.g., if
        the initial mass function were top-heavy. Studies have shown, however, that
        this function should be the same at all redshifts $z\lesssim 8$
        \cite{Dias2010}. Hopefully, this conclusion can be tested using
        supernova rates in the future, which may eliminate even this last
        possible caveat for the significant tension between the observed
        halo mass function and $\Lambda$CDM. On the flip side, if it turns
        out that future observations with JWST support an evolution in the halo to stellar
        mass ratio of at least $\sim 0.8$ dex between $z\sim 4$ and $10$, validating the
        predictions of $\Lambda$CDM, the inferred halo distribution will be in tension
        with the predictions of $R_{\rm h}=ct$. The differences are so significant (at least
        several orders of magnitude) that a refinement of the halo distribution may produce
        one of the most robust comparative cosmological tests of these models.

        \section{Conclusion}
        In this paper, we have discussed an ongoing debate concerning the
        early appearance of massive galaxies (and their halos), which may challenge the formation
        of structure predicted by $\Lambda$CDM if the halo to stellar mass ratio is roughly constant
        in the redshift range $4<z<10$. This difficulty could be mitigated with a refinement of the
        underlying theory of star formation and galaxy evolution, but appears to require implausible
        modifications to the physics underlying these phenomena (Steinhardt et al. 2016). Some
        support for the existence of a real problem is provided by other types of ``too early''
        problems, such as the premature appearance of supermassive black holes at $z\sim 6-7$
        \cite{Melia2013b,MeliaMcClintock2015}.

        Combining our earlier measurement of $\sigma_8$ at redshift 0 \cite{Melia2017}
        with our recently completed calculation of the growth function using the coupled Boltzmann
        and perturbed Einstein equations, we have re-analyzed
        ``The Impossibly Early Galaxy Problem" in the context of $R_{\rm h}=ct$
        and showed that this problem virtually disappears in this cosmology
        even if the halo to stellar mass ratio is constant. Although, the $R_{\rm h}=ct$
        universe overpredicts the number density of halos by one to two orders of magnitude at the very
        high mass end, this problem may be mitigated by corrections to the Sheth-Tormen mass function,
        as indicated by the Bolshoi simulations \cite{Klypin2011}. Thus, once we resolve the
        question of whether or not this ratio evolved with redshift, the inferred halo
        mass distribution can clearly distinguish between the $R_{\rm h}=ct$ and
        $\Lambda$CDM cosmologies.

        The timeline in $R_{\rm h}=ct$ allows both massive galaxies and
        supermassive black holes to form at very high redshifts without invoking
        exotic physics. It should also be noted that, while $\Lambda$CDM
        must rely on the unproven and as yet unverified physics of inflation
        to account for the generation of scale-invariant primordial fluctuations
        and a mechanism for driving the modes to exit and re-enter the horizon, thus
        creating an intricate mechanism for producing different growth rates at
        different epochs, no such complicated, fine-tuned mechanism is necessary
        in $R_{\rm h}=ct$. This model does not have a horizon problem and does
        not incorporate inflation into its expansion history. As explained in
        more detail in ref.~\cite{YennapureddyMelia2019}, the growth of structure
        in $R_{\rm h}=ct$ is simple, streamlined and does not require a different
        handling of small modes compared to the larger ones. Such simplicity,
        particularly when viewed in the context of the excellent agreement between
        theory and observations (figs.~2-7), adds considerable support for the
        viability of this cosmology.

        Looking forward to upcoming surveys and further theoretical developments,
        it is already clear that observations, e.g., with JWST, will play a crucial
        role in determining the quasar distribution and the rate of gamma ray
        bursts from Pop III stars, both heavily dependent on the growth rates
        we have been discussing in this paper. There is therefore significant
        promise of improving the comparison we have made here even further, perhaps
        strongly ruling out one or other of these two models.

{\acknowledgement
        We are grateful to Charles Steinhardt and Peter Behroozi
        for very informative discussions. FM is grateful to the Instituto de Astrof\'isica de
        Canarias in Tenerife and to Purple Mountain Observatory in Nanjing, China for their
        hospitality while part of this research was carried out.
\endacknowledgement}

%
%

\end{document}